\documentclass[prl,twocolumn,letterpaper, superscriptaddress]{revtex4-1}
\usepackage{graphicx}
\usepackage{dcolumn}
\usepackage{bm}
\usepackage{color}
\usepackage{tabularx}
\usepackage{array}
\usepackage{amsmath}
\usepackage{pbox}
\usepackage{stmaryrd}  

\begin{document}

\title{Tunable magnetically induced transparency spectra in magnon-magnon coupled Y$_3$Fe$_5$O$_{12}$/permalloy bilayers}

\author{Yuzan Xiong}
\affiliation{Department of Physics, Oakland University, Rochester, MI 48309, USA}
\affiliation{Department of Electronic and Computer Engineering, Oakland University, Rochester, MI 48309, USA}
\affiliation{Materials Science Division, Argonne National Laboratory, Argonne, IL 60439, USA}

\author{Jerad Inman}
\affiliation{Department of Physics, Oakland University, Rochester, MI 48309, USA}
\affiliation{Department of Electronic and Computer Engineering, Oakland University, Rochester, MI 48309, USA}

\author{Zhengyi Li}
\affiliation{Jiangsu Key Laboratory of Opto-Electronic Technology, School of Physics and Technology, Nanjing Normal University, Nanjing 210046, China}

\author{Kaile Xie}
\affiliation{Jiangsu Key Laboratory of Opto-Electronic Technology, School of Physics and Technology, Nanjing Normal University, Nanjing 210046, China}

\author{Rao Bidthanapally}
\affiliation{Department of Physics, Oakland University, Rochester, MI 48309, USA}

\author{Joseph Sklenar}
\affiliation{Department of Physics and Astronomy, Wayne State University, Detroit, MI 48201, USA}

\author{Peng Li}
\affiliation{Department of Electrical and Computer Engineering, Auburn University, Auburn, AL 36849, USA}

\author{Steven Louis}
\affiliation{Department of Electronic and Computer Engineering, Oakland University, Rochester, MI 48309, USA}

\author{Vasyl Tyberkevych}
\affiliation{Department of Physics, Oakland University, Rochester, MI 48309, USA}

\author{Hongwei Qu}
\affiliation{Department of Electronic and Computer Engineering, Oakland University, Rochester, MI 48309, USA}

\author{Zhili Xiao}
\affiliation{Materials Science Division, Argonne National Laboratory, Argonne, IL 60439, USA}

\author{Wai K. Kwok}
\affiliation{Materials Science Division, Argonne National Laboratory, Argonne, IL 60439, USA}

\author{Valentine Novosad}
\affiliation{Materials Science Division, Argonne National Laboratory, Argonne, IL 60439, USA}

\author{Yi Li}
\thanks{yili@anl.gov}
\affiliation{Materials Science Division, Argonne National Laboratory, Argonne, IL 60439, USA}

\author{Fusheng Ma}
\thanks{phymafs@njnu.edu.cn}
\affiliation{Jiangsu Key Laboratory of Opto-Electronic Technology, School of Physics and Technology, Nanjing Normal University, Nanjing 210046, China}

\author{Wei Zhang}
\thanks{weizhang@oakland.edu}
\affiliation{Department of Physics, Oakland University, Rochester, MI 48309, USA}

\date{\today}

\begin{abstract}

Hybrid magnonic systems host a variety of characteristic quantum phenomena such as the magnetically-induced transparency (MIT) and Purcell effect, which are considered useful for future coherent quantum information processing. In this work, we experimentally demonstrate a tunable MIT effect in the Y$_3$Fe$_5$O$_{12}$(YIG)/Permalloy(Py) magnon-magnon coupled system via changing the magnetic field orientations. By probing the magneto-optic effects of Py and YIG, we identify clear features of MIT spectra induced by the mode hybridization between the uniform mode of Py and the perpendicular standing spin wave modes of YIG. By changing the external magnetic field orientations, we observe a tunable coupling strength between the YIG's spin-wave modes and the Py's uniform mode, upon the application of an out-of-plane magnetic field. This observation is theoretically interpreted by a geometrical consideration of the Py and YIG magnetization under the oblique magnetic field even at a constant interfacial exchange coupling. Our findings show high promise for investigating tunable coherent phenomena with hybrid magnonic platforms.  

\end{abstract}

\maketitle

\section{I. Introduction}

Hybrid magnonic systems are rising contenders for coherent information processing, owing to their capability of coherently connecting distinct physical platforms in quantum systems and their rich emerging quantum engineering functionalities \cite{nakamura_apex2019,hu_ssp2018,bhoi_ssp2019,tqe_2021,jap_2021}. Recent studies have revealed strong, coherent hybridization of magnons with phonons, microwave photons, and optical light, with the observation of characteristic phenomena such as the strong and ultra-strong coupling, magnetically-induced transparency (MIT), and Purcell effect \cite{nakamura_science2015,haigh_prb2019,yili_prl2019,luqiao_prl2019,huebl_prl2013,bai_prl2015,xufeng_prl2014,xufeng_ncomm2015}. In addition, to fully leverage the hybrid coupling phenomena with magnons, strong and tunable couplings between two magnonic systems have recently attracted considerable attention \cite{weiler_prl2018,yu_prl2018,qin_srep2018,yili_magnon2019,xiong_srep2020}. These systems can be considered as hosting hybrid magnonic modes in a ``magnonic cavity", in analogous to the microwave photonic cavity in cavity-magnon polaritons.

\begin{figure*}[htb]
 \centering
 \includegraphics[width=6.75 in]{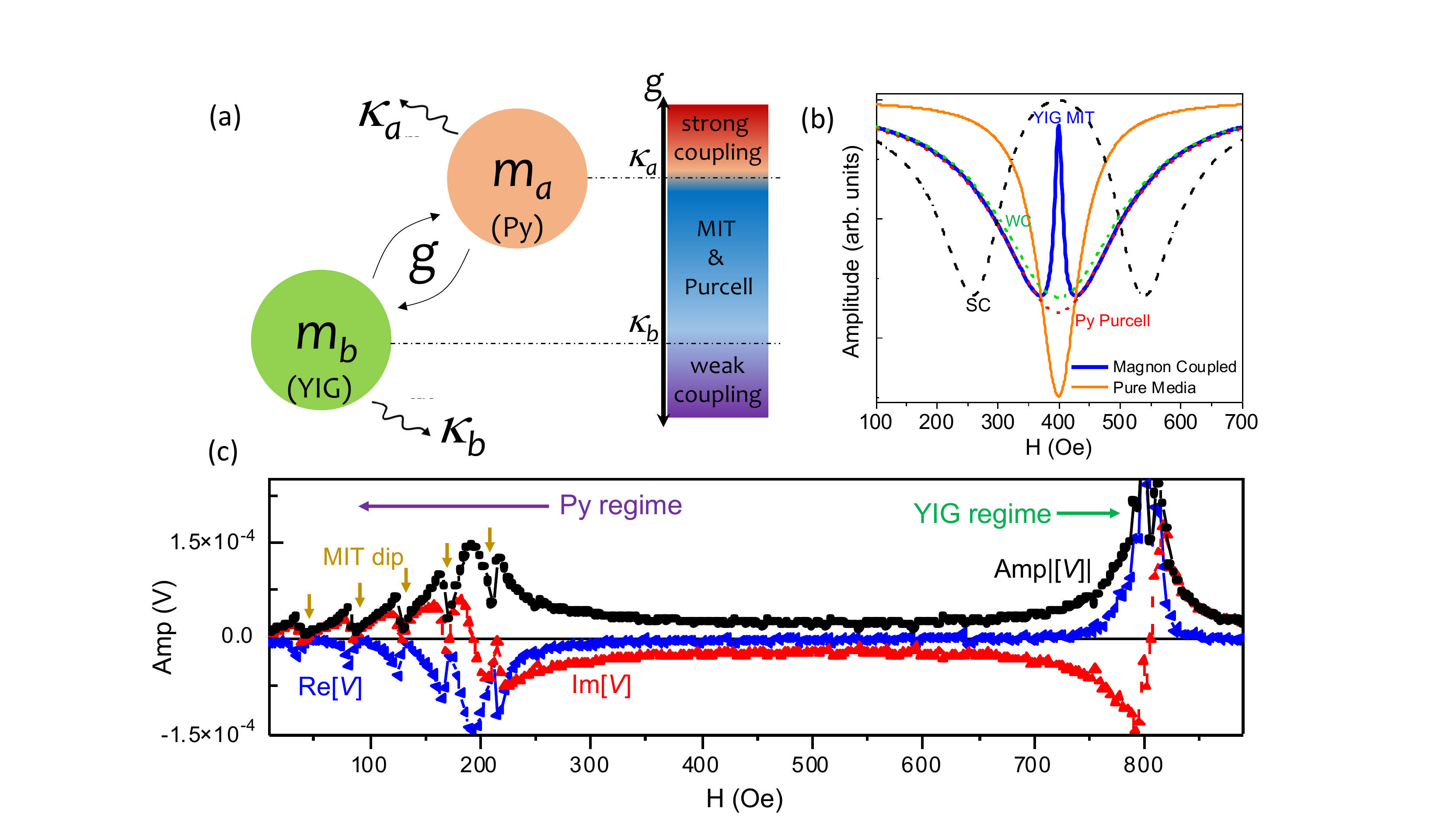}
 \caption{(a) Schematics of the linearly coupled magnon-magnon system: $m_a$(Py) $\rightleftharpoons$ $m_b$(YIG). $g$, $\kappa_a$, $\kappa_b$ are the coupling strength and dissipation rates of the respective magnon subsystems. Different coupling regimes are characterized and separated by the relative strength between the coupling rate and dissipation rates of the two magnon subsystems. (b) Illustration of the coupled mode spectra in the MIT/Purcell regime ($\kappa_b < g < \kappa_a$). In the field domain, the mode hybridization leads to an abrupt suppression of the absorption at a certain field window. Such a transparency window, whose bandwidth is determined by the low-loss mode (YIG), can be observed in the broad resonance of the other lossy mode (Py).  Reciprocally, the lossy mode exhibits enhanced decay due to the Purcell effect. Such resonant phenomena can be controlled by an external magnetic field. As a comparison, the modeled trace for the case of strong coupling (SC) and weak coupling (WC) are also plotted. (c) An example magneto-optical signal trace of YIG/Py(30-nm) showing the lineshapes of the in-phase $X$ (Re[$V$]) and quadrature $Y$ (Im[$V$]), and the total amplitude, Amp[$|V|$]$=\sqrt{X^2+Y^2}$, measured at 4-GHz at $\theta = 0^{\circ}$, $\phi = 0^{\circ}$ ($h \perp H$). \textcolor{black}{The low-field ($0 - 300$ Oe) regime corresponds to the Py-FMR regime, where the hybridized YIG PSSW modes strongly modulate the Py resonance profile, exhibiting the ``MIT dip" (vertical arrows). The high-field ($700 - 900$ Oe) regime corresponds to the YIG-FMR regime.}}
 \label{fig1}
\end{figure*}

\begin{figure*}[htb]
 \centering
 \includegraphics[width=4.85 in]{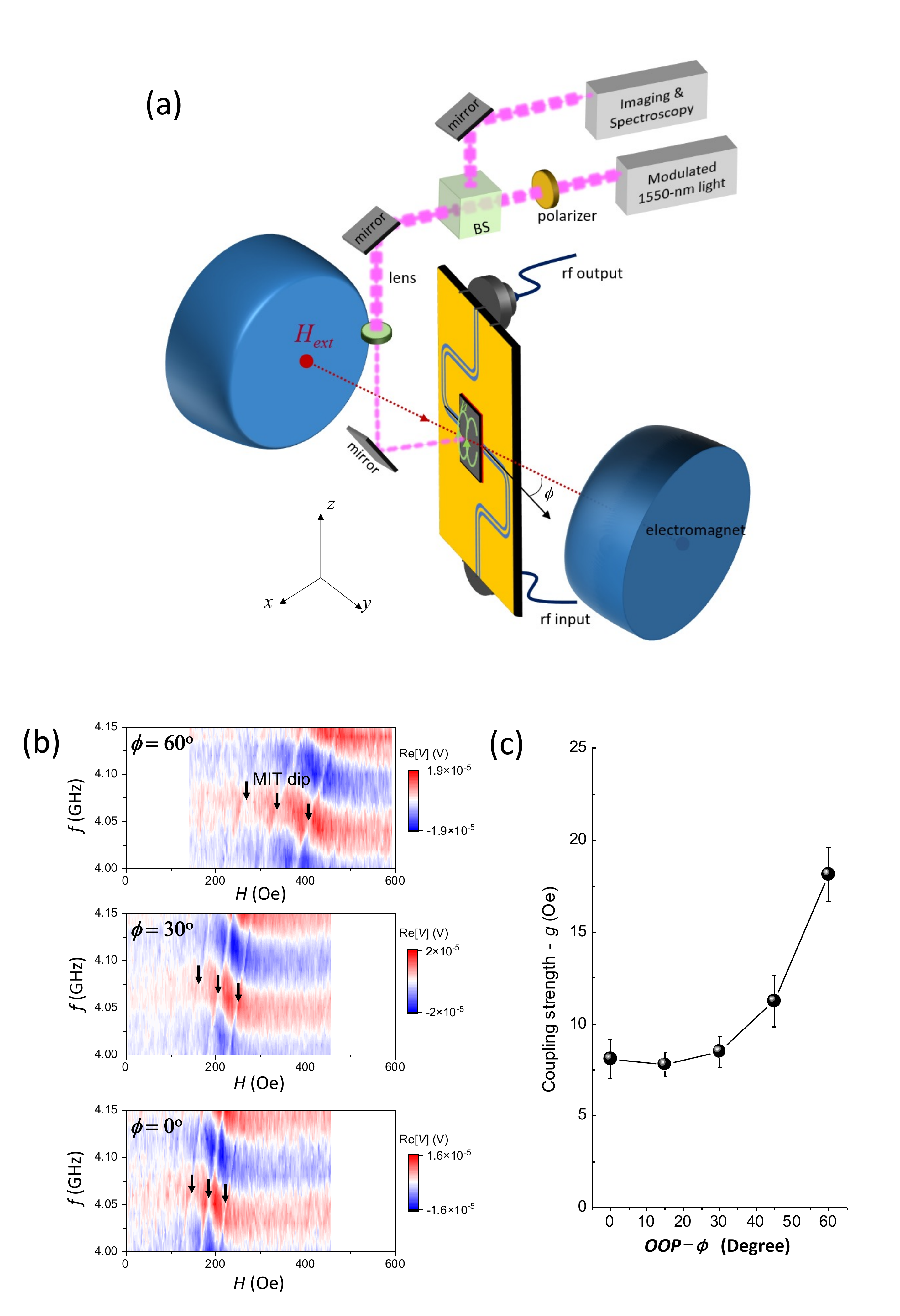}
 \caption{(a) Schematics of the experimental magneto-optic setup for OOP magnetic field configuration. A 1550-nm light is sourced and modulated by fiber-optic components before entering the real-space. Key components in the real-space includes a polarizer, a beam-splitter (BS), a focus lens, and mirrors for detouring the light path. The microwave excitation for the sample is achieved by a coplanar waveguide in a flip-chip configuration. Light passes through the YIG layer and is retro-reflected against the opaque Py layer, before it is directed towards a balancing photodiode and subsequent signal analysis. (b) Example contour plots of the measured Re[$V(f,H)$] at $\phi =$ 0$^\circ$, 30$^\circ$, and 60$^\circ$, \textcolor{black}{where the MIT dips are marked by the arrows}. (c) OOP$-\phi$ angular dependence of the extracted coupling strength $g$.  }
 \label{fig2}
\end{figure*}

So far, studies of magnon-magnon coupling have been largely centered around the strong coupling regime, with a characteristic observation of the anti-crossing gap between the two intersecting magnonic modes. In such a strong coupling regime, the coupling strength, $g/2\pi$, is greater than the dissipation rate of both coupling entities, $\kappa_a/2\pi$, $\kappa_b/2\pi$, as illustrated in Fig. \ref{fig1}(a). On the other hand, the regime where $g/2\pi$ is weaker than both dissipation rates is known as the weak coupling. In between the strong and weak coupling regimes, the coupling satisfies $\kappa_b < g < \kappa_a$, and two characteristic quantum phenomena can be observed, namely, the MIT and Purcell effects \cite{xufeng_prl2014}, depending on which magnon resonator to be pumped. The MIT effect is the magnetic analogy of electromagnetically-induced transparency (EIT). One example is a spin-wave induced suppression of the ferromagnetic resonance (FMR) of another adjacent layer \cite{poimanov_jphysd2019,poimanov_prb2018}. The Purcell effect is characterized with an enhanced decay of the cavity photon due to its coupling to lossy magnons. In conventional magnon-photon coupled systems, the two effects are respectively discussed with the magnon and the photon subsystems. In magnon-magnon coupled systems, the two effects are reciprocal, i.e. the MIT effect of $m_a$ is simultaneously accompanied with the Purcell effect of $m_b$, as illustrated in Fig. \ref{fig1}(b). Such MIT effect in magnon-magnon coupled systems was earlier reported in Y$_3$Fe$_5$O$_{12}$(YIG)/Permalloy(Py) thin-film bilayers, by probing the coupled magnetization dynamics via the magneto-optic Kerr and Faraday effects \cite{xiong_srep2020}.

\section{II. Sample and Measurement}

In this work, we further examine the magnetic field tunability of the magnon-magnon coupled mode spectra in the MIT regime. We probe the resonant magneto-optical signals of YIG/Py bilayers under two different magnetic field varying configurations, i.e., out-of-plane(OOP)-$\phi$ and in-plane(IP)-$\theta$ rotating field geometries. In addition, our measurement setup allows to optionally introduce a static OOP field by a permanent magnet when the magnetic field is swept in-plane. A detailed depiction of the measurement construction is included in the Supplemental Material (SM) \cite{sm}. 

Our YIG/Py bilayers were fabricated by magnetron-sputtering Py thin films on 3-$\mu$m-thick, single-sided, commercial YIG films grown on double-side-polished Gd$_3$Ga$_5$O$_{12}$ (GGG) substrates using liquid phase epitaxy (LPE). Following earlier recipes for ensuring a good YIG/Py interfacial exchange coupling \cite{xiong_srep2020}, we used \textit{in-situ} Argon gas rf-bias cleaning for 3 min in the vacuum chamber, to clean the YIG surface before depositing a 30-nm-Py layer. \textcolor{black}{In the earlier report \cite{xiong_srep2020}, we also fabricated and measured the YIG/SiO$_2$/Py and YIG/Cu/Py reference samples. The SiO$_2$ and the Cu layers can both effectively break the interfacial exchange coupling, thus resulting in a null observation of the magnon-magnon coupled characteristics.}   

The samples were chip-flipped on a coplanar waveguide (CPW) for broadband microwave excitation. The experimental details of spin-dynamic excitation and magneto-optic detection can be found in our earlier report \cite{xiong_srep2020}. We use a heterodyne technique in which a 1550-nm laser light is linearly-polarized and modulated at the FMR frequencies simultaneously with the sample's excitation \cite{yili_prap2019,yili_ieee2019,xiong_ieee2021, ono_apl2021}, and detect the dynamic Faraday and Kerr signals upon the excitation of Py-FMR and YIG's perpendicular standing spin waves (PSSWs). Due to the dispersion relation of YIG and Py, the Py-FMR mode couples to the YIG-PSSW modes and effectively serves as a "lossy cavity". 

Due to the limited spatial overlap, the YIG-PSSW modes couple weakly to the CPW's microwave drive. However, as shown by a typical trace in Fig. \ref{fig1}(c) measured at 4 GHz, through the interfacial exchange coupling, the Py-FMR mode can efficiently excite the PSSW modes under its envelope ($H \sim 50 - 250$ Oe), and resonantly enhancing the YIG dynamics. \textcolor{black}{At the same time, the YIG PSSW modes strongly modulate the Py resonance profile, exhibiting the ``MIT dip".} The magneto-optic signal, $V$, with the phase information, are obtained by the lock-in's in-phase $X$ (Re[$V$]) and quadrature $Y$ (Im[$V$]), which are further used to calculate the total amplitude, Amp[$|V|$] $=\sqrt{X^2+Y^2}$. By considering a series of YIG harmonic oscillators coupled with the Py oscillator, the coupled spin dynamics in the MIT regime can be phenomenologically modeled, and the measured complex magneto-optic signal, $V$, can be expressed as: 

\begin{widetext}
    \begin{equation}
        V = \frac{A e^{i(\phi_\text{L}-\phi_\text{m})}}{i(H^\text{Py}_\text{FMR} - H) - \Delta H_\text{Py} + \Sigma_n\frac{g^2}{i(H^\text{YIG}_{\text{PSSW},n} - H) - \Delta H_{\text{YIG},n}}}
    \label{eq01}
    \end{equation}
\end{widetext}
where $A$ is the total signal amplitude, $H_\text{FMR}^\mathrm{Py}$ and $H_\text{PSSW}^\mathrm{YIG}$ is the resonance field of Py and YIG-PSSWs, respectively, $\Delta H_\mathrm{YIG(Py)}$ is the half-width-half-maximum linewidth, $n$ is the PSSW mode-index number, and $g$ is the fieldlike coupling strength from the interfacial exchange. According to our previous reports \cite{yili_magnon2019,xiong_srep2020}, the YIG and Py are antiferromagnetically coupled at the interface. $\phi_{L}$ is the phase accumulated due to the optical path length, and $\phi_m$ is the magnetization phase. More details regarding the physical contributions to $\phi_{L}$ and $\phi_{m}$ can be found in the earlier reports \cite{xiong_srep2020,xiong_ieee2021}. To examine the magnetic field tunability of the magnon-magnon coupled spectra, we performed measurements on the YIG/Py(30-nm) sample under three experimental configurations: OOP rotating field, IP rotating field, and IP rotating field with a static OOP disturbing field by using a permanent magnet. All experiments were performed at room temperature. 

\begin{figure*}[htb]
 \centering
 \includegraphics[width=6.8 in]{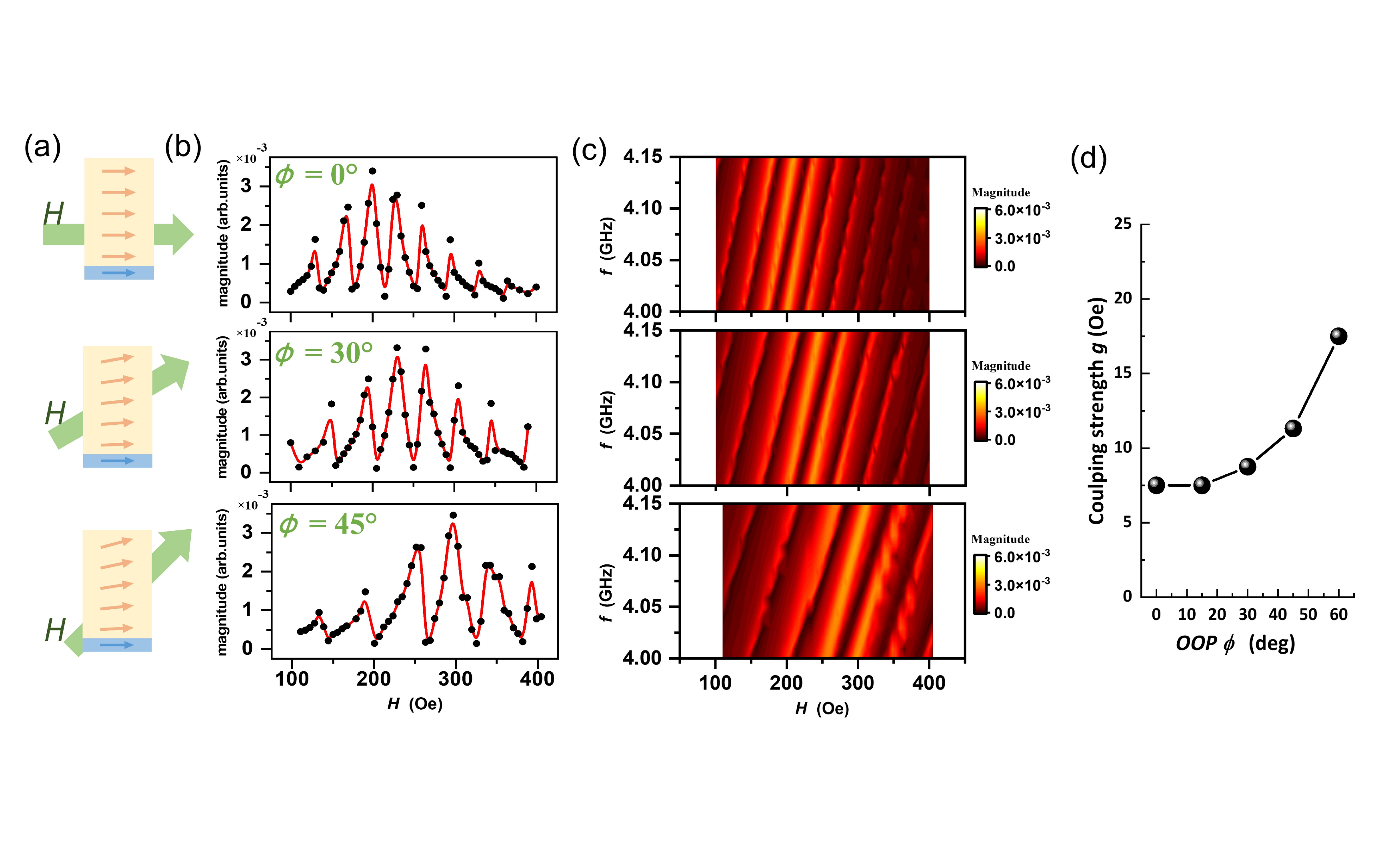}
 \caption{(a) Illustration of the oblique magnetic field effect: the blue/yellow areas represent the Py/YIG layer with arrows indicate the direction of magnetization. Green arrows indicate the direction of the external magnetic field $H$. (b) Simulated microwave absorption spectrum at 4.1 GHz for representative OOP$-\phi$ angles: $\phi =$ 0$^\circ$, 30$^\circ$, and 45$^\circ$, \textcolor{black}{reproducing the experimentally observed MIT dips}. (c) Corresponding color contour plots of the simulated $f-H$ dispersion from 4.0 - 4.15 GHz: $\phi =$ 0$^\circ$, 30$^\circ$, and 45$^\circ$. (d) Simulated dependence of the coupling strength $g$ on the OOP-$\phi$ angle. }
 \label{fig3}
\end{figure*}

\section{III. Results and Discussions}
\subsection{A. Out-of-plane magnetic field dependence}

\textcolor{black}{Earlier experimental reports have shown an enhanced magnon-magnon coupling due to the introduction of out-of-plane magnetic field, magnetic anisotropy, and Dzyaloshinskii–Moriya interaction (DMI) in magnetic multilayers, particularly in the synthetic antiferromagnets (SAFs) systems \cite{macneill_prl2019,shiota_prl2020,sud_prb2020,sklenar_prap2021}. It is theoretically explained by the additional symmetry breaking due to the out-of-plane effective field that increases the magnon-magnon interaction \cite{he_prb2021}. Beyond the SAFs where the magnetic layers are coupled by the Ruderman–Kittel–Kasuya–Yosida (RKKY) interaction, it is also worthwhile to investigate similar effects in the direct exchange-coupled YIG/Py, which has been a prototypical magnon-magnon system. In addition, since the YIG and Py has distinct magnetic characteristics, it would be important to examine how the magnetization geometry of the two layers influences their magnon-magnon coupled spin dynamics. }

Our measurement setup and dataset for the OOP field experiment are shown in Fig.\ref{fig2}. The external magnetic field generated by the electromagnet can be rotated out-of-plane ($x$-$y$), at an angle $\phi$ with respect to the sample plane ($y$-$z$), as illustrated in Fig. \ref{fig2}(a). The Oersted field, $h$, supplied by the CPW, is along $z$. We scanned the frequency, $f$, and magnetic field, $H$, around the Py-FMR (MIT regime), and plotted the Re[$V(f,H)$] map at incremental $\phi$ angles from 0$^\circ$ to 60$^\circ$, at a step of 15$^\circ$. Figure \ref{fig2}(b) shows the contour plots of the MIT spectra at 0$^\circ$, 30$^\circ$, and 60$^\circ$. The two-dimensional plots are generated from the individual Re[$V$] spectra acquired by sweeping the magnetic field at a given frequency (from 4 to 4.15 GHz). From the contour plots, it can be seen that the Py-FMR is strongly modulated by a series of sharp YIG PSSW modes, \textcolor{black}{showing a series of ``MIT dip"}. A group of 5 $\sim$ 6 PSSWs can be clearly identified under the Py uniform excitation. Due to the much lower resonance field of Py compared to YIG, see Fig.\ref{fig1}(c), these PSSWs are usually higher-index modes, typically with an index number, $n \sim 30 - 40$, with the corresponding wavenumbers $k=n\pi/d_{YIG} \sim 31.5 - 42.0$ $\mu$m$^{-1}$, where $d_{YIG}$ is the YIG thickness, according to our previous report \cite{xiong_srep2020}.

To analyze the coupling strength $g$, we have chosen the two to three PSSW modes that are close to the center of the Py resonance envelope, and fit the experimental $V(H)$ traces with Eq.\ref{eq01}. The coupling $g$, and the dissipation rates for YIG and Py are subsequently extracted. In the field domain, the lower limit of the field linewidth for the lossy magnon cavity of Py is $\sim 30.0$ Oe, and the upper limit of the YIG linewidth is $\sim 2.0$ Oe. On the other hand, the extracted coupling $g$, as plotted in Fig.\ref{fig2}(c), is in the range of $\sim 7.0 - 20.0$ Oe. Therefore, the hybrid system is confirmed in the MIT coupling regime, satisfying $\kappa_{YIG} < g < \kappa_{Py}$. Notably, the coupling strength $g$ increases with increasing OOP angle, $\phi$. At $\phi = 60^\circ$, the coupling is more than doubled than the IP case ($\phi = 0^\circ$). Such an angular dependence is in a general agreement with the magnon coupled acoustic and optic modes \textcolor{black}{in the SAFs reported earlier, which also features an antiferromagnetic coupling at the interface, but via the RKKY interaction} \cite{macneill_prl2019,shiota_prl2020,sud_prb2020,sklenar_prap2021}. 

In order to verify the tunable coupling between the PSSW modes of YIG and the FMR mode of Py, micromagnetic simulations were performed to numerically solve the Landau-Lifshitz-Gilbert equation by using MuMax3 \cite{mumax3}, with the effective magnetic field including contributions from the interlayer exchange field, the intralayer exchange field, the demagnetization field, and the external field. The strength of the interlayer exchange interaction between the YIG and the Py layers is determined by the exchange energy density $J$. The modeled thickness of the Py and YIG layers are the same as in the experiment: $t_{Py} =$ 30 nm and $t_{YIG} =$ 3000 nm, respectively. The whole Py/YIG bilayer structure with size 200 × 200 × 3030 nm$^3$ is discretized into 10 × 10 × 1212 cells. The periodic boundary conditions are applied in the in-plane two directions along the $y-z$. The material parameters used in simulations are as: for Py layer, $M_s^{Py}$ = 8.6 × 10$^5$ A/m, $A_{ex}^{Py}$ = 1.3 × 10$^{-11}$ J/m, and $\alpha_{Py}$ = 0.01; for YIG layer, $M_s^{YIG}$ = 1.4 × 10$^5$ A/m, $A_{ex}^{YIG}$ = 3.5 × 10$^{-12}$ J/m, and $\alpha_{YIG}$ = 5 × 10$^{-4}$. The strength of interlayer exchange interaction is determined by $J$ = 0.3 × 10$^{-3}$ J/m$^2$. 

To excite PSSW modes in the YIG layer, a driving field $h_z(t)$ is locally applied in the bottommost layer of the Py along the $z$-axis \cite{zhizhi_prb2021}. The $h_z(t)$ is in the form of “sinc” function \cite{apl_ma2011}: $h_z(t)=h_0$sin$[2 \pi f (t- t_0)]/2 \pi f (t- t_0)$, with the cut-off frequency $f$ = 50 GHz, the offset time $t_0=$ 0.25 ns, and the amplitude $h_0$ = 1 mT. The spatially averaged magnetization of all cells is recorded every 10 ps during the total simulation time of 100 ns. The spin-wave spectra can be obtained by Fourier transforming the out-of-plane component of the magnetization in the Py layer. 

Figure \ref{fig3}(a) illustrates the oblique magnetic field effect to the magnetization distribution of the YIG/Py bilayer. The simulated microwave absorption spectra and the $f-H$ dispersion curves are shown in Fig. \ref{fig3}(b) and (c), respectively, for representative OOP-angles, $\phi =$ 0$^\circ$, 30$^\circ$, and 45$^\circ$. It is found that the splitting of the hybridized modes between the uniform mode of Py and the PSSW modes of YIG is enhanced by increasing the $\phi$ angle. To better understand the splitting, the coupling strength $g$ is extracted from the microwave absorption spectra according to Eq.\ref{eq01}. The simulated $g$ value is 7.51 Oe, 8.75 Oe, and 11.31 Oe, at $\phi =$ 0$^\circ$, 30$^\circ$, and 45$^\circ$, respectively. Figure \ref{fig3}(d) shows that the $g$ increases with increasing $\phi$ from 0$^\circ$ to 60$^\circ$, which is consistent with the experimental observation shown in Fig.\ref{fig2}. 

Such a dependence can be understood by the geometrical factor of YIG and Py macrospins as well as the external field induced non-collinear magnetization distributions along the thickness direction of the YIG/Py bilayer as shown in Fig.\ref{fig3}(a). When the external field $H$ is applied out-of-plane, the magnetization in Py is estimated always in the plane in the studied field range. However, the magnetization of YIG along the thickness direction is spatially dependent on the distance to Py and gradually tilting towards $H$. The further away from the Py, the magnetization in the YIG is closer to the direction of $H$. The larger the difference between the magnetization of Py and YIG, the larger the coupling strength $g$, which results from the $H$-induced spatial symmetry breaking. Similar observations have been also reported in the magnetic bilayer of Co wires deposited on YIG film \cite{yu_prl2018}. 

\begin{figure*}[htb]
 \centering
 \includegraphics[width=5.0 in]{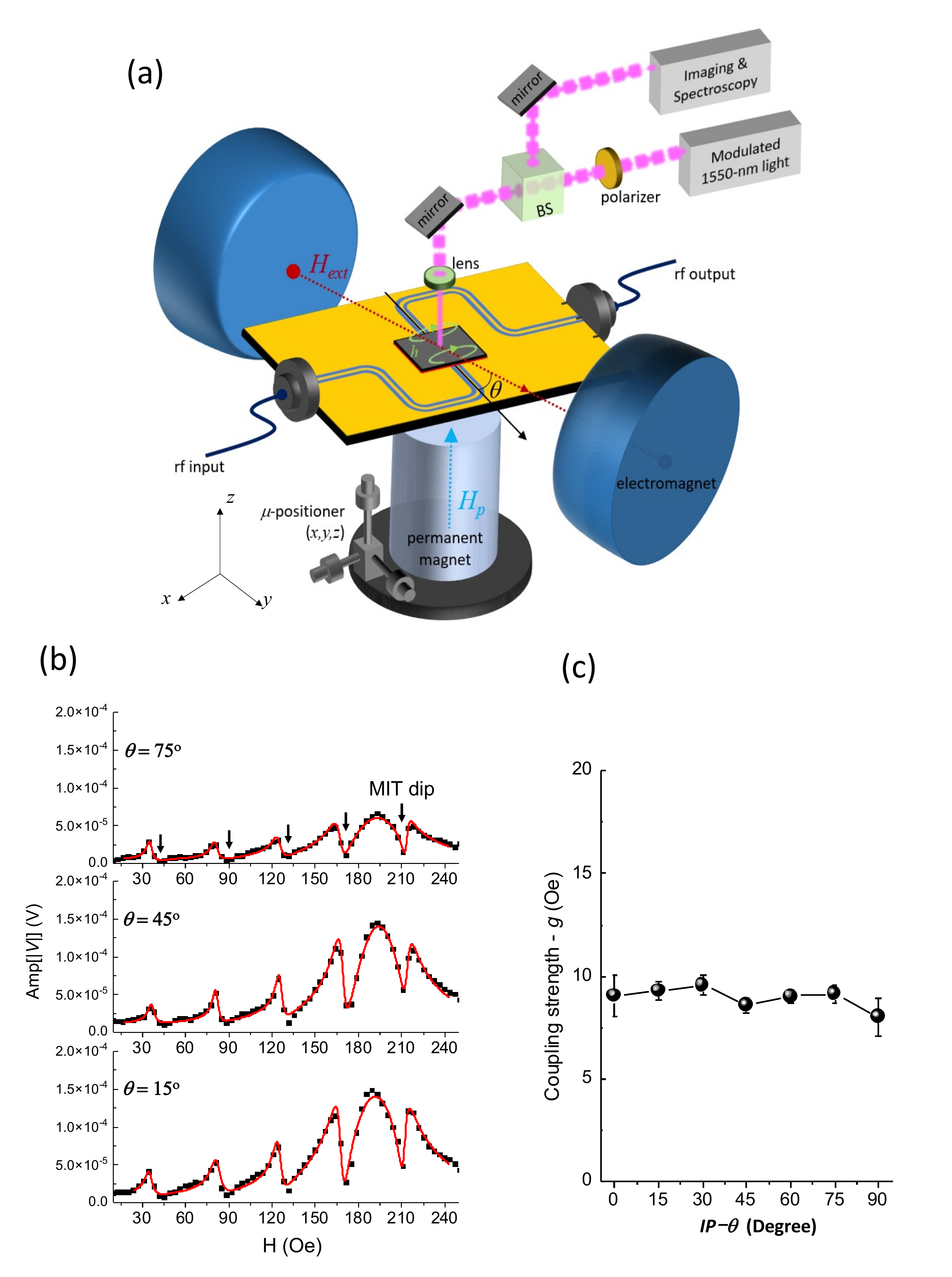}
 \caption{(a) Schematics of the experimental magneto-optic setup for IP magnetic field configuration (with optional OOP disturbing fields). (b) Example signal traces of $V(H)$ measured at 4-GHz (dots) and the corresponding fitting curves (lines) following the Eq.\ref{eq01}, at $\theta =$ 15$^\circ$, 45$^\circ$, and 75$^\circ$. \textcolor{black}{A total of 5 MIT dips can be observed (arrows). As the $\theta$ increases, the amplitude of the Py-FMR and the MIT dip decrease at approximately the same rate, therefore leading to a negligible dependence of $g$ versus $\theta$. (c) The extracted IP$-\theta$ angular dependence of the extracted coupling strength $g$.} }
 \label{fig4}
\end{figure*}

From a theory perspective, a comprehensive analytical model would be difficult to develop due to the many factors that may change with increasing the OOP $\phi-$angle, such as the mode ellipticities, excitation of PSSWs with different mode index $n$, and the dipolar interactions \cite{sklenar_prap2021,jeffrey_apl2021,fusheng_apl2021,bozhko_prr2020,yu_prl2018}. On the other hand, our earlier work \cite{xiong_srep2020} and the subsequent sections (with in-plane angle dependence) suggest that the coupling strength $g$ is a weak function of the PSSW mode number $n$ at higher indices, $n>25$. In addition, the contribution to the coupling by the change of the precessional ellipticity is estimated to be less than $\sim 3\%$. However, one important factor that has not been well discussed so far is the additional dipolar interaction that may be caused by the obliquely applied magnetic field. To this end, we have theoretically modeled the effect of such dipolar interaction to the coupling between the Py and YIG spin-wave mode, which can be found in the SM \cite{sm}. 

By solving the matrix element of the dipolar interaction calculated for complex spin-wave profiles of the interacting modes, we found an additional, dipolar coupling strength, $g_{dip}$, that depends on the magnetization directions, due to the geometrical restriction of the excited spin wave modes. A theoretical relationship between the coupling and the internal OOP magnetization angles of Py ($\phi_{Py}$) and YIG ($\phi_{YIG}$) can be derived as: 
\begin{equation}
g_{dip} = g_0+ \kappa_0 h(\phi_{YIG},\phi_{Py}) 
\label{eq02}
\end{equation}
where $g_0$ and $\kappa_0$ are geometrical-independent constants (Details are included in the SM \cite{sm}). The geometrical function, $h(\phi_{YIG},\phi_{Py})$, can be expressed as: $h(\phi_{YIG},\phi_{Py}) = 1 - \textrm{cos}(\phi_{YIG})\textrm{cos}(\phi_{Py}) + \frac{1}{2}\textrm{sin}(\phi_{YIG})\textrm{sin}(\phi_{Py})$, and this  angular function $h(\phi_{YIG},\phi_{Py})$ is an increasing function of $\phi_{YIG}$. On the other hand, due to a much higher magnetization in the Py layer, the OOP magnetization angle of Py, $\phi_{Py}$, will be smaller than $\phi_{YIG}$. Therefore, the OOP oblique field to the YIG/Py bilayer should result in a net enhancement of the dipolar coupling between the YIG's spin wave modes and the Py-FMR. A detailed theoretical description can be found in the SM \cite{sm}.

\subsection{B. In-plane magnetic field dependence}

Next, we discuss the measurement setup and dataset for the IP field experiment. \textcolor{black}{Generally, for the PSSWs that are of interest here, one would not expect any in-plane field dependence due to the fact that the wavenumber is pointing out-of-plane, and both the YIG and Py are sufficiently soft. However, due to our CPW-excitation, the rf field is fixed along $x$ (in the IP configuration), it would be useful then to examine and verify any possible dependence that may be caused by the relative orientation between the rf field and the magnetization. }

Our measurement setup and dataset for the IP field experiment are summarized in Fig.\ref{fig4}. In this configuration, the sweeping magnetic field supplied by the electromagnet can be rotated in-plane ($x$-$y$), at an angle $\theta$, with respect to the central CPW signal line (along $y$), see Fig.\ref{fig4}(a). The Oersted field, $h$, supplied by the CPW, is along $x$. Similarly, we scanned the frequency and magnetic field around the MIT regime,  and at incremental $\theta$ angles from 0$^\circ$ to 90$^\circ$, at a step of 15$^\circ$. In such an IP configuration, we can also optionally introduce an additional OOP disturbing field, $H_p$, from the back of the sample, via an axially-magnetized permanent magnet placed on a micro-positioner, to further modify the MIT spectra (results to be discussed later).

Figure \ref{fig4}(b) shows example $V(H)$ traces at 4-GHz measured at $\theta = 15^\circ, 45^\circ$, and 75$^\circ$ around the Py-FMR regime. We again analyze the coupling strength $g$ via fitting the hybridized PSSW modes near the center of the Py-FMR in the $V(H)$ curves with Eq.\ref{eq01}. The extracted coupling, $g$, as plotted in Fig.\ref{fig4}(c), is $\sim 7.0 - 10.0$ Oe, and has indicated a negligible IP-$\theta$ angular dependence, \textcolor{black}{despite a global signal amplitude attenuation as $\theta$ increases}. Since the interfacial exchange (the dominant driving force) has a negligible anisotropy and both YIG and Py are soft magnets, the anisotropic excitation caused by the CPW signal-line plays a negligible role in the magnon-magnon hybridization. In addition, with both the YIG and Py macrospins aligned in-plane, the influence from the geometrical factor and the possibly associated dipolar interaction do not contribute to additional coupling terms. 

We note that earlier reports in dealing with SAFs have suggested a nontrivial IP angular dependence along with a wavenumber dependence of the coupling strength $g$ \cite{macneill_prl2019,shiota_prl2020,sud_prb2020}. Here, although using different $g$ values for the $V(H)$ fitting can in principle lead to an improved fitting result, we do not find a convincing $k$-dependent coupling $g$, \textcolor{black}{which is also expected for the hybridized PSSW modes. This is because that the wavenumber of the PSSWs is primarily pointing out-of-plane, and the dipolar coupling between the PSSWs and the in-plane Py are sufficiently weak \cite{zhizhi_prb2021}. Besides, due to the large difference in the magnetization of YIG and Py, the PSSW modes that are actively coupled to the Py uniform mode are of higher-index ($n \sim 30 - 40$), and their dipolar field is much less than the uniform mode and even those long wavelength (smaller $n$) PSSW modes \cite{xiong_srep2020}.} However, this does not exclude the possible contribution of dipolar enhanced coupling when the magnetization are tilted out-of-plane, as we discussed in the OOP case earlier.  

\begin{figure*}[htb]
 \centering
 \includegraphics[width=7 in]{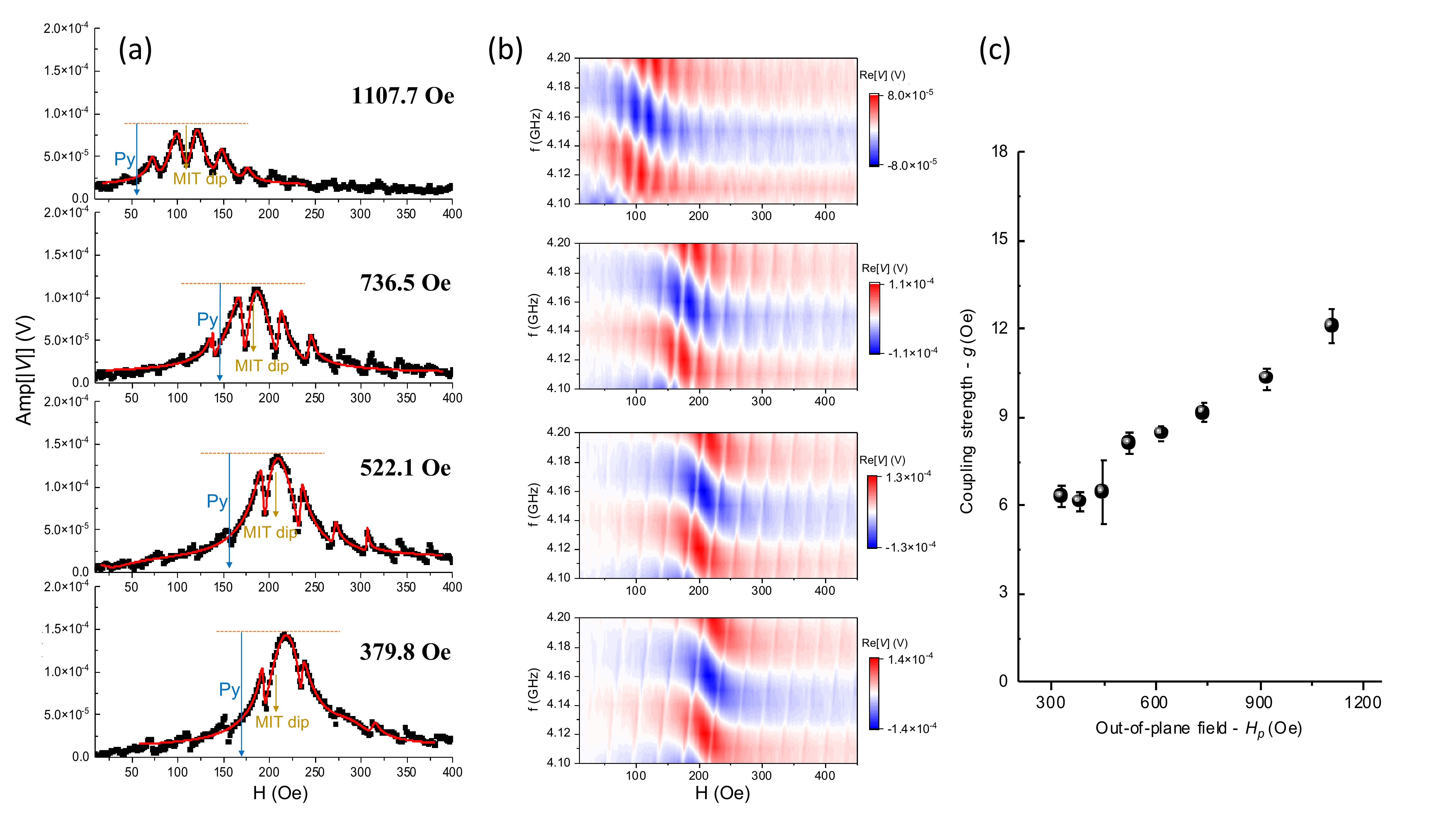}
 \caption{(a) Example $V(H)$ signal traces measured at 4.16-GHz (dots) and the corresponding fitting curves (lines) following the Eq.\ref{eq01}, and (b) corresponding Re[$V(f,H)$] contour plots, at selective $H_p$ values: 379.8, 522.1, 736.5, and 1107.7 Oe. \textcolor{black}{In (a), the modulation effect to the Py-FMR profile can be visualized by directly comparing the total Py amplitude and the modulation depth (the MIT dip) caused by the hybrid YIG-PSSW modes (arrows).} (c) The $H_p-$dependence of the extracted coupling strength $g$. }
 \label{fig5}
\end{figure*}

\subsection{C. Out-of-plane bias magnetic field}

We then discuss the effect of introducing an OOP disturbing field, $H_p$ on the MIT spectra and the coupling $g$ in the IP measurement configuration. \textcolor{black}{Different from the OOP field sweeping discussed earlier in Sec. A, our experiment here examines how a static OOP bias field, mimicking effectively an unidirectional anisotropy, may influences the magnon-magnon coupled characteristics. In realworld material systems, such an effective field may be provided by either a magnetic anisotropy or a DMI field.}

The disturbing field is provided by a large-diameter (0.5-inch dia.), axially-magnetized, permanent magnet that is fixed on a micro-positioner, see Fig.\ref{fig4}(a), and placed in close proximity with the sample/CPW stack from the backside. Such a static field serves as a constant ``bias field" to the same field-sweep experiment described in Fig.\ref{fig4}. By adjusting the position of the magnet (primarily along $z$), the magnitude of $H_p$ can be tuned. After first performing a calibration of the magnetic field, we found a uniform-field range of $\sim 350 - 1200$ Oe, which corresponds to a sample-magnet gap distance in the range of $\sim 0.5 - 7.0$ mm. We performed similar measurements in the IP configuration under such a static disturbing field at a fixed $\theta = 0^\circ$, and the results are summarized in Fig.\ref{fig5}. 

Figure \ref{fig5}(a) shows example $V(H)$ traces measured at 4.16-GHz at selective $H_p$ values, 379.8, 522.1, 736.5, and 1107.7 Oe. Increasing the magnitude of $H_p$ leads to an overall shift of the MIT regime towards lower field range, due to the modification to the Py-FMR dispersion. In addition, the magnon-magnon coupling becomes stronger as $H_p$ increases, which is evidenced by the enhanced modulation effect to the Py-FMR profile caused by the hybrid PSSW modes. \textcolor{black}{Such a modulation effect can be visualized simply by comparing the total Py amplitude and the modulation depth (the MIT dip) caused by the hybridization}. Figure \ref{fig5}(b) shows the corresponding contour plots of Re[$V(f,H)$] at the same $H_p$ values as in Fig.\ref{fig5}(a). The frequency range is scanned around the MIT regime, from 4.1 to 4.2 GHz, and the field is swept from 0 to 450 Oe. 

Similar to the above OOP and IP cases, we can also analyze the MIT spectra under such a bias OOP field $H_p$, and extract a coupling strength $g$ via fitting the hybridized PSSW modes near the center of the Py-FMR in the $V(H)$ curves with Eq.\ref{eq01}. The extracted coupling, $g$ is further plotted in Fig.\ref{fig5}(c). For $H_p$ values below $\sim 500.0$ Oe, no significant effect can be found. However, for $H_p$ values above $\sim 500.0$ Oe, a robust enhancement of the coupling $g$ is observed, which steadily increases as $H_p$ increases. This result echos our earlier experiment with the OOP configuration. The enhanced coupling $g$ could be again likewise attributed to the similar \textcolor{black}{symmetry breaking and the geometrical consideration of YIG and Py macrospins}, as well as the enhanced dipolar interactions due to the tilted magnetization. Finally, we note that our experiment with the OOP disturbing field also reflects a tunable signal response and a great spectrum sensitivity at the MIT regime, which may be found useful in magnetic sensing applications with magnon-magnon coupled systems.

\section{V. Conclusion}

In summary, we experimentally study the magnetic field tunability of the magnetically-induced transparency spectra in YIG/Py magnon-magnon coupled bilayers. By probing the magneto-optic effects of Py and YIG, we identify clear features of MIT spectra induced by the mode hybridization between the FMR of Py and the PSSWs of YIG. By changing the external magnetic field orientation out-of-plane, we observe a tunable coupling strength between the YIG's spin-wave modes and the Py's uniform mode. Such a tunable coupling can be understood by a geometrical consideration of the oblique magnetization vectors of the Py and YIG, which enhances the coupling via an additional coupling effect from the dipolar interaction. On the other hand, rotating the magnetic field in-plane results in no observable changes to the magnon-magnon coupling strength, which agrees with the standing wave nature of the excited spin wave modes. Further, by introducing a static bias magnetic field using a permanent magnet, we reconfirmed such an geometrical effect to the coupling strength under an out-of-plane magnetic field component. Our reported experiments may be found useful for future studies on the tunable coherent phenomena with hybrid magnonic platforms.

\textcolor{black}{ \section*{Acknowledgment}
The experimental measurement and theoretical modeling at Oakland University was supported by the U.S. National Science Foundation under Grant No. ECCS-1941426. The sample fabrication and preparation at Argonne National Laboratory was supported by  U.S. DOE, Office of Science, Materials Sciences and Engineering Division. The micromagnetic simulations at Nanjing Normal University was supported by the National Natural Science Foundation of China (Grant No. 12074189). Z.-L.X. acknowledges funding support from the U.S. National Science Foundation under Grant No. DMR-1901843. We thank Dr. Vivek Amin and Zhizhi Zhang for fruitful discussions and \textcolor{black}{Mr. Lei Zhang and Dr. Cassie Ward for structural and surface characterizations. Part of the characterization also made use of the PXDR facility that is partially funded by NSF under Grant No. MRI-1427926 at Wayne State University.} }



\begin{thebibliography}{0}%
\makeatletter
\providecommand \@ifxundefined [1]{%
 \@ifx{#1\undefined}
}%
\providecommand \@ifnum [1]{%
 \ifnum #1\expandafter \@firstoftwo
 \else \expandafter \@secondoftwo
 \fi
}%
\providecommand \@ifx [1]{%
 \ifx #1\expandafter \@firstoftwo
 \else \expandafter \@secondoftwo
 \fi
}%
\providecommand \natexlab [1]{#1}%
\providecommand \enquote  [1]{``#1''}%
\providecommand \bibnamefont  [1]{#1}%
\providecommand \bibfnamefont [1]{#1}%
\providecommand \citenamefont [1]{#1}%
\providecommand \href@noop [0]{\@secondoftwo}%
\providecommand \href [0]{\begingroup \@sanitize@url \@href}%
\providecommand \@href[1]{\@@startlink{#1}\@@href}%
\providecommand \@@href[1]{\endgroup#1\@@endlink}%
\providecommand \@sanitize@url [0]{\catcode `\\12\catcode `\$12\catcode
  `\&12\catcode `\#12\catcode `\^12\catcode `\_12\catcode `\%12\relax}%
\providecommand \@@startlink[1]{}%
\providecommand \@@endlink[0]{}%
\providecommand \url  [0]{\begingroup\@sanitize@url \@url }%
\providecommand \@url [1]{\endgroup\@href {#1}{\urlprefix }}%
\providecommand \urlprefix  [0]{URL }%
\providecommand \Eprint [0]{\href }%
\providecommand \doibase [0]{http://dx.doi.org/}%
\providecommand \selectlanguage [0]{\@gobble}%
\providecommand \bibinfo  [0]{\@secondoftwo}%
\providecommand \bibfield  [0]{\@secondoftwo}%
\providecommand \translation [1]{[#1]}%
\providecommand \BibitemOpen [0]{}%
\providecommand \bibitemStop [0]{}%
\providecommand \bibitemNoStop [0]{.\EOS\space}%
\providecommand \EOS [0]{\spacefactor3000\relax}%
\providecommand \BibitemShut  [1]{\csname bibitem#1\endcsname}%
\let\auto@bib@innerbib\@empty
\end{thebibliography}%


\begin{thebibliography}{19}

\bibitem{nakamura_apex2019} D. Lachance-Quirion, Y. Tabuchi, A. Gloppe, K. Usami, and Y. Nakamura, ``Hybrid quantum systems based on magnonics", Applied Physics Express \textbf{12}, 070101 (2019).

\bibitem{hu_ssp2018} M. Harder and C. -M. Hu, ``Cavity Spintronics: An Early Review of Recent Progress in the Study of Magnon-Photon Level Repulsion”, Solid State Physics, \textbf{70}, 47 - 121 (2018). R. Stamps and R. Camley (Ed.), Academic Press.

\bibitem{bhoi_ssp2019} B. Bhoi and S. -K. Kim, ``Photon-magnon coupling: Historical perspective, status, and future directions”, Solid State Physics, \textbf{69}, 1 - 77 (2019). R. Stamps and H. Schultheiss (Ed.), Academic Press.

\bibitem{tqe_2021} D. D. Awschalom, C.H.R. Du, R. He, J. Heremans, A. Hoffmann, J. Hou, H. Kurebayashi, Y. Li, L. Liu, V. Novosad, J. Sklenar, S. Sullivan, D. Sun, H. Tang, V. Tyberkevych, C. Trevillian, A. W. Tsen, L. Weiss, W. Zhang, X. Zhang, L. Zhao, and Ch. W. Zollitsch, ``Quantum Engineering With Hybrid Magnonics Systems and Materials", IEEE Trans. Quantum Engineering \textbf{2}, 5500836 (2021).

\bibitem{jap_2021} Y. Li, W. Zhang, V. Tyberkevych, W.-K. Kwok, A. Hoffmann, V. Novosad, ``Hybrid magnonics: Physics, circuits, and applications for coherent information processing", J. Appl. Phys. \textbf{128}, 130902 (2020).

\bibitem{nakamura_science2015} Y. Tabuchi, S. Ishino, A. Noguchi, T. Ishikawa, R. Yamazaki, K. Usami, and Y. Nakamura, ``Coherent coupling between a ferromagnetic magnon and a superconducting qubit", Science \textbf{349}, 405 (2015).

\bibitem{haigh_prb2019} L. McKenzie-Sell, J. Xie, C.-M. Lee, J. W. A. Robinson, C. Ciccarelli, and J. A. Haigh, ``Low-impedance superconducting microwave resonators for strong coupling to small magnetic mode volumes", Phys. Rev. B \textbf{99}, 140414 (2019).

\bibitem{yili_prl2019} Y. Li, T. Polakovic, Y.-L. Wang, J. Xu, S. Lendinez, Z. Zhang, J. Ding, T. Khaire, H. Saglam, R. Divan, J. Pearson, W. K. Kwok, Z. Xiao, V. Novosad, A. Hoffmann, and W. Zhang, ``Strong Coupling between Magnons and Microwave Photons in On-Chip Ferromagnet-Superconductor Thin-Film Devices", Phys. Rev. Lett. \textbf{123}, 107701 (2019).

\bibitem{luqiao_prl2019} J. T. Hou and L. Liu, ``Strong coupling between microwave photons and nanomagnet magnons", Phys. Rev. Lett. \textbf{123}, 107702 (2019).

\bibitem{huebl_prl2013} H. Huebl, C. W. Zollitsch, J. Lotze, F. Hocke, M. Greifenstein, A. Marx, R. Gross, and S. T. B. Goennenwein, ``High Cooperativity in Coupled Microwave Resonator Ferrimagnetic Insulator Hybrids", Phys. Rev. Lett. \textbf{111}, 127003 (2013).

\bibitem{bai_prl2015} L. Bai, M. Harder, Y. P. Chen, X. Fan, J. Q. Xiao, and C. M. Hu, ``Spin Pumping in Electrodynamically Coupled Magnon-Photon Systems", Phys. Rev. Lett. \textbf{114}, 227201 (2015).

\bibitem{xufeng_prl2014} X. Zhang, C.-L. Zou, L. Jiang, and H. X. Tang, ``Strongly Coupled Magnons and Cavity Microwave Photons", Phys. Rev. Lett. \textbf{113}, 156401 (2014).

\bibitem{xufeng_ncomm2015} X. Zhang, C.-L. Zou, N. Zhu, F. Marquardt, L. Jiang, and H. X. Tang, ``Magnon dark modes and gradient memory", Nat. Commun. \textbf{6}, 8914 (2015).

\bibitem{weiler_prl2018} S. Klingler, V. Amin, S. Gepr¨ags, K. Ganzhorn, H. Maier-Flaig, M. Althammer, H. Huebl, R. Gross, R. D. McMichael, M. D. Stiles, S. T. B. Goennenwein, and M. Weiler, ``Spin-Torque Excitation of Perpendicular Standing Spin Waves in Coupled YIG/Co Heterostructures", Phys. Rev. Lett. \textbf{120}, 127201 (2018).

\bibitem{yu_prl2018} J. Chen, C. Liu, T. Liu, Y. Xiao, K. Xia, G. E. W. Bauer, M. Wu, and H. Yu, ``Strong Interlayer Magnon-Magnon Coupling in Magnetic Metal-Insulator Hybrid Nanostructures", Phys. Rev. Lett. \textbf{120}, 217202 (2018).

\bibitem{qin_srep2018} H. Qin, S. J. Hamalainen, and S. van Dijken, ``Exchange-torque-induced excitation of perpendicular standing spin waves in nanometer-thick YIG films", Sci. Rep. \textbf{8}, 5755 (2018).

\bibitem{yili_magnon2019} Y. Li, W. Cao, V. P. Amin, Z. Zhang, J. Gibbons, J. Sklenar, J. Pearson, P. M. Haney, M. D. Stiles, W. E. Bailey, V. Novosad, A. Hoffmann, and W. Zhang, ``Coherent spin pumping in a strongly coupled magnon-magnon hybrid system", Phys. Rev. Lett. \textbf{124}, 117202 (2020).

\bibitem{xiong_srep2020} Y. Xiong, Y. Li, M. Hammami, R. Bidthanapally, J. Sklenar, X. Zhang, H. Qu, G. Srinivasan, J. Pearson, A. Hoffmann, V. Novosad, and W. Zhang, "Probing magnon–magnon coupling in exchange coupled Y3Fe5O12/Permalloy bilayers with magneto-optical effects", NPG Sci. Rep. \textbf{10}, 12548 (2020). 

\bibitem{poimanov_jphysd2019} V. D. Poimanov, A. N. Kuchko, and V. V. Kruglyak, ``Emission of coherent spin waves from a magnetic layer excited by a uniform microwave magnetic field", J. Phys. D: Appl. Phys. \textbf{52}, 135001 (2019). 

\bibitem{poimanov_prb2018} V. D. Poimanov, A. N. Kuchko, and V. V. Kruglyak, ``Magnetic interfaces as sources of coherent spin waves", Phys. Rev. B \textbf{98}, 104418 (2018).

\bibitem{sm} See Supplemental Material at http://link.aps.org/supplemental/XXXXX for additional details on the experimental setup, theoretical analysis, and supporting sample characterizations. 

\bibitem{yili_prap2019} Y. Li, H. Saglam, Z. Zhang, R. Bidthanapally, Y. Xiong, J. E. Pearson, V. Novosad, H. Qu, G. Srinivasan, A. Hoffmann, and W. Zhang, ``Simultaneous Optical and Electrical Spin-Torque Magnetometry with Phase-Sensitive Detection of Spin Precession", Phys. Rev. Applied \textbf{11}, 034047 (2019).

\bibitem{yili_ieee2019} Y. Li, F. Zeng, H. Saglam, J. Sklenar, J. E. Pearson, T. Sebastian, Y. Wu, A. Hoffmann, and W. Zhang, ``Optical Detection of Phase-Resolved Ferromagnetic Resonance in Epitaxial FeCo Thin Films", IEEE Trans. Magn. \textbf{55}, 6100605 (2019).

\bibitem{xiong_ieee2021} Y. Xiong, Y. Li, R. Bidthanapally, J. Sklenar, M. Hammami, S. Hall, X. Zhang, P. Li, J. E. Pearson, T. Sebastian, G. Srinivasan, A. Hoffmann, H. Qu, V. Novosad, and W. Zhang ``Detecting Phase-Resolved Magnetization Dynamics by Magneto-Optic Effects at 1550 nm Wavelength", IEEE Trans. Magn. \textbf{57}, 4300807 (2021). 

\bibitem{ono_apl2021}  Y. Shiota,  R. Hisatomi,  T. Moriyama,  A. S. Samardak, and  T. Ono, ``Inhomogeneous magnetic properties characterized by simultaneous electrical and optical detection of spin-torque ferromagnetic resonance", Appl. Phys. Lett. \textbf{119}, 192409 (2021). 

\bibitem{macneill_prl2019} D. MacNeill, J. T. Hou, D. R. Klein, P. Zhang, P. Jarillo-Herrero, and L. Liu, ``Gigahertz Frequency Antiferromagnetic Resonance and Strong Magnon-Magnon Coupling in the Layered Crystal CrCl3", Phys. Rev. Lett. \textbf{123}, 047204 (2019). 

\bibitem{shiota_prl2020} Y. Shiota, T. Taniguchi, M. Ishibashi, T. Moriyama, and T. Ono, ``Tunable Magnon-Magnon Coupling Mediated by Dynamic Dipolar Interaction in Synthetic Antiferromagnets", Phys. Rev. Lett. \textbf{125}, 017203 (2020). 

\bibitem{sud_prb2020} A. Sud, C. W. Zollitsch, A. Kamimaki, T. Dion, S. Khan, S. Iihama, S. Mizukami, and H. Kurebayashi, ``Tunable magnon-magnon coupling in synthetic antiferromagnets", Phys. Rev. B \textbf{102}, 100403(R) (2020). 

\bibitem{sklenar_prap2021} J. Sklenar and W. Zhang, ``Self-Hybridization and Tunable Magnon-Magnon Coupling in van der Waals Synthetic Magnets", Phys. Rev. Applied \textbf{15}, 044008 (2021). 

\bibitem{he_prb2021} M. Li, J. Lu, and W. He, ``Symmetry breaking induced magnon-magnon coupling in synthetic antiferromagnets", Phys. Rev. B \textbf{103}, 064429 (2021). 

\bibitem{mumax3} A. Vansteenkiste, J. Leliaert, M. Dvornik, M. Helsen, F. Garcia-Sanchez, and B. Van Waeyenberge, ``The Design and Verification of MuMax3", AIP Adv. \textbf{4}, 107133 (2014).

\bibitem{zhizhi_prb2021} Z. Zhang, H. Yang, Z. Wang, Y. Cao, and P. Yan, ``Strong coupling of quantized spin waves in ferromagnetic bilayers",  Phys. Rev. B \textbf{103}, 104420 (2021). 

\bibitem{apl_ma2011} F. S. Ma, H. S. Lim, Z. K. Wang, S. N. Piramanayagam, S. C. Ng, and M. H. Kuok, ``Micromagnetic Study of Spin Wave Propagation in Bicomponent Magnonic Crystal Waveguides", Appl. Phys. Lett. \textbf{98}, 153107 (2011).


\bibitem{jeffrey_apl2021} T. Jeffrey, W. Zhang, and J. Sklenar, ``Effect of dipolar interaction on exceptional points in synthetic layered magnets", Appl. Phys. Lett. \textbf{118}, 202401 (2021). 

\bibitem{fusheng_apl2021} C. Dai and F. Ma, ``Strong magnon–magnon coupling in synthetic antiferromagnets", Appl. Phys. Lett. \textbf{118}, 112405 (2021).  

\bibitem{bozhko_prr2020} D. A. Bozhko, H. Yu. Musiienko-Shmarova, V. S. Tiberkevich, A. N. Slavin, I. I. Syvorotka, B. Hillebrands, and A. A. Serga, ``Unconventional spin currents in magnetic films", Phys. Rev. Research \textbf{2}, 023324 (2020). 





\end{thebibliography}
\end{document}